\pdfoutput=1
\documentclass[a4paper,UKenglish]{lipics-v2018}

\usepackage{microtype}
\usepackage{makeidx}  
\usepackage{programs}
\usepackage{graphicx}
\graphicspath{{images/}}

\title{Concurrent Robin Hood Hashing}

\author{Robert Kelly}{Maynooth University Department of Computer Science, Maynooth, Ireland}{rob.kelly@cs.nuim.ie}{https://orcid.org/0000-0001-8266-2961}{}

\author{Barak A. Pearlmutter}{Maynooth University Department of Computer Science and Hamilton Institute, Maynooth, Ireland}{barak@cs.nuim.ie}{https://orcid.org/0000-0003-0521-4553}{}

\author{Phil Maguire}{Maynooth University Department of Computer Science, Maynooth, Ireland}{pmaguire@cs.nuim.ie}{https://orcid.org/0000-0002-8993-8403}{}

\authorrunning{R. Kelly, B. A. Pearlmutter, P. Maguire}

\Copyright{Robert Kelly,  Barak A. Pearlmutter, and Phil Maguire}

\subjclass{}


\category{}

\relatedversion{}

\supplement{}

\funding{Funded by the Government of Ireland Postgraduate Scholarship.}

\acknowledgements{I want to thank William Leiserson for his invaluable help with reviewing and feedback. I also want to thank Nir Shavit for extensive access to hardware for running the benchmarks.}

\EventEditors{Jiannong Cao, Faith Ellen, Luis Rodrigues, and Bernardo Ferreira}
\EventNoEds{4}
\EventLongTitle{22nd International Conference on Principles of Distributed Systems (OPODIS 2018)}
\EventShortTitle{OPODIS 2018}
\EventAcronym{OPODIS}
\EventYear{2018}
\EventDate{December 17–19, 2018}
\EventLocation{Hong Kong, China}
\EventLogo{}
\SeriesVolume{122}
\ArticleNo{0} 
\nolinenumbers 
\hideLIPIcs  

\begin{document}

\maketitle

\begin{abstract}
In this paper we examine the issues involved in adding concurrency to the Robin Hood hash table algorithm. We present a non-blocking obstruction-free \textit{K-CAS} Robin Hood algorithm which requires only a single word compare-and-swap primitive, thus making it highly portable. The implementation maintains the attractive properties of the original Robin Hood structure, such as a low expected probe length, capability to operate effectively under a high load factor and good cache locality, all of which are essential for high performance on modern computer architectures. We compare our data-structures to various other lock-free and concurrent algorithms, as well as a simple hardware transactional variant, and show that our implementation performs better across a number of contexts.
\end{abstract}

\section{Introduction}
Concurrent data-structures allow multiple threads to operate on them without risk of data corruption, as well as providing guarantees of correctness for concurrent operations. Lock-free data-structures are a class of concurrent data-structures that have specific properties relating to system or thread progress guarantees. The programming of portable and practical lock-free data-structures is becoming ever more practical, with the addition of mainstream language support for atomic variables, and a well defined thread memory model (see \cite{C++11}, \cite{JavaMemModel}).

Concurrent algorithms can be separated into two major classes, namely blocking and non-blocking, with both featuring further partitions based on the specific progress guarantees within those classes \cite{ampp}. Blocking algorithms have well documented issues when it comes to their use. They are susceptible to deadlock, priority inversion, convoying, and a lack of composability with respect to multiple operations on data-structures. A lock-free, or non-blocking, algorithm has none of these problems. Such algorithms suffer, however, from their own set of challenges relating to memory management (\cite{Alistarh:2015:TAS:2755573.2755600}, \cite{Alistarh:2017:FCM:3064176.3064214}, \cite{Michael:2004:HPS:987524.987595}, \cite{fraser2004practical}, \cite{cohen2015efficient}), correctness (\cite{Herlihy:1990:LCC:78969.78972}, \cite{Leino2009}, \cite{DBLP:conf/pdp/AmighiBH16}) and potentially lackluster performance as the system is flooded with contention under heavy write load. As hardware manufacturers resort to expanding processor core counts for enhanced performance \cite{sutterlunch}, non-blocking data-structures are coming to the fore, providing more robust progress guarantees and tolerance to the suspension of threads. For end consumers to realise the full performance of their system, algorithms must efficiently exploit as many cores as possible.

Hash tables are one of the major building blocks in software applications, providing efficient implementations for the abstract data types of maps and sets. These data-structures are highly versatile, making them an active area of research in concurrency (e.g. \cite{Michael2002}, \cite{ampp}, \cite{Hopper}, \cite{CliffClick}, \cite{Nielsen:2016:SLH:3016078.2851196}, \cite{Purcell2005}, \cite{Shalev03}). Hash tables are associative data-structures that contain a pool of keys and associated values \cite{Cormen:2001:IA:580470}, lending themselves to efficient implementations. In general, they feature the methods \texttt{Add}, \texttt{Remove}, and \texttt{Contains}, each of which is bound by $\mathcal{O}(1)$ computational complexity while requiring $\mathcal{O}(n)$ space \cite{Cormen:2001:IA:580470}. Hash table algorithms achieve this performance by calculating an index, called a \textit{hash}, from each key, and use this to efficiently find the relevant entry in the pool of keys, normally an array. Unlike comparative structures, which store keys in a sorted order and binary search through the space, hash tables rely on the hash function to distribute the keys across the space. Ideally, the hash function generates a unique index for each key. In reality, however, the keys often have the same hash, creating what is known as a \textit{collision}. A primary focus of research in hash tables is how to efficiently deal with these collisions.

Hash tables can be divided into two major design variants, namely open-addressing and closed-addressing (i.e. separate-chaining). Open addressing stores the key and value pairings in different buckets in the table, either through a pointer or directly in the internal array itself, with a single item allowed per bucket. When a hash collision takes place (when another entry has taken the desired bucket), a new bucket is selected via some collision resolution algorithm. Separate chaining, on the other hand, stores a pointer to a list of values at that bucket, containing all the key and value pairings that collided on that bucket.

Robin Hood Hashing \cite{Celis:1986:RHH:13298} is an open-addressing hash table method in which entries in the table are moved around so that the variance of the distances to their original bucket is minimised. Insertion is a multi-stage process, potentially moving multiple items throughout the table. The general goal is to find an empty bucket, or another entry that is less `deserving' of its bucket than the current item being inserted or moved. If an empty bucket is found, it is taken. If another less deserving entry has been identified, then it's swapped with the current entry and `kicked' further down the line, with this process repeating until an empty bucket is found. The serial version of Robin Hood is well suited to modern computer architecture. As CPU utilisation effectively becomes a function of the memory bottleneck \cite{drepper07memory}, algorithms that use the CPU cache more efficiently can enhance performance for memory bounded tasks. For instance, Robin Hood Hashing has a very low expected probe count, allowing reads to be culled early even though the algorithm uses linear probing. Low probe counts mean fewer cache misses, performing very well on modern architectures. As it stands, these attractive properties have been maintained in our concurrent versions. Our concurrent solution manages to achieve non-blocking progress  with physical deletion and all of the aforementioned benefits of the serial algorithm. We use an algorithmically optimised \textit{K-CAS} \cite{ArbelRaviv2017ReuseDR} implementation along with a timestamp mechanism to handle bulk relocations while maintaining correctness.

In Section 2 we review the current landscape of concurrent and lock-free hash tables, and the original Robin Hood algorithm. Section 3 outlines the structure of our algorithms and the various challenges encountered in adding concurrency, while Section 4 discusses the performance of these algorithms relative to competitors.

\section{Background}

\subsection{Prior Work}
A number of concurrent open-addressing hash table algorithms have been proposed. Purcell and Harris \cite{Purcell2005} presented a lock-free open-addressing hash table where per-bucket upper bounds are stored in conjunction with the keys, thus allowing searches to be culled early. Nielson and Karlsson \cite{Nielsen:2016:SLH:3016078.2851196} built upon the work of Purcell and Harris with a Lock-Free Linear Probing hash table, simplifying the earlier algorithm, reducing the number of bucket states required, and removing the word normally required. Herlihy, Shavit, and Tzafrir \cite{Hopper} presented Hopscotch Hashing, a concurrent hash table algorithm with outstanding performance. This algorithm allows searches, insertions, and removals to skip over irrelevant items, and is also cache aware in its reordering of entries present in the table. While Hopscotch's insertions and deletions are blocking, the mutating operations are \textit{sharded} over multiple locks.

Like open-addressing, separate-chaining can be implemented in many different forms. Michael \cite{Michael2002} presented a lock-free hash table, with each bucket containing a slightly modified Harris linked list \cite{Harris:2001:PIN:645958.676105}. Shalev and Shavit \cite{Shalev03} presented a particularly succinct implementation of using a linked list as a hash table, indexing into it from an array of node pointers. In this case, the list can grow forever, so long as pointers to new entry points are added to prevent the probe length from growing out of control. The table can be automatically resized, as only the entry point array of pointers need be extended. Laborder, Feldman, and Dechev \cite{Feldman2015} presented their Wait-Free Hash Table, whereby a key collision at a bucket leads the bucket to be expanded into another sub-table until a point is reached such that all collisions are resolved.

\subsection{Original Robin Hood}
Robin Hood Hashing was first proposed by Celis \cite{Celis:1986:RHH:13298} in 1986. Though remaining relatively obscure, it has recently gained recognition via the new programming language Rust \cite{RustHashTable}, which has adopted Robin Hood as its standard hash table algorithm. Robin Hood is an open-addressing hash table algorithm that employs linear probing for finding items and for finding spaces for new entries. Robin Hood does exactly what it says on the tin: it steals from the rich and gives to the poor. Here, ``rich'' refers to items that got lucky during the hashing process, by finding a free bucket close to their original bucket. In other words, these ``rich'' items have a low \textbf{D}istance \textbf{F}rom their expected \textbf{B}ucket (this distance is referred to as \textit{DFB} for short) and a low expected probe count before being found. In contrast, being ``poor'' means hashing to a bucket that has been heavily saturated beforehand, and thus has a high number of items to step over before finding a free bucket. ``Poor'' items thus have a higher \textit{DFB}, and a high expected probe count before being found.

\begin{figure*}[ht]
  \centering
  \includegraphics[width=0.95\textwidth]{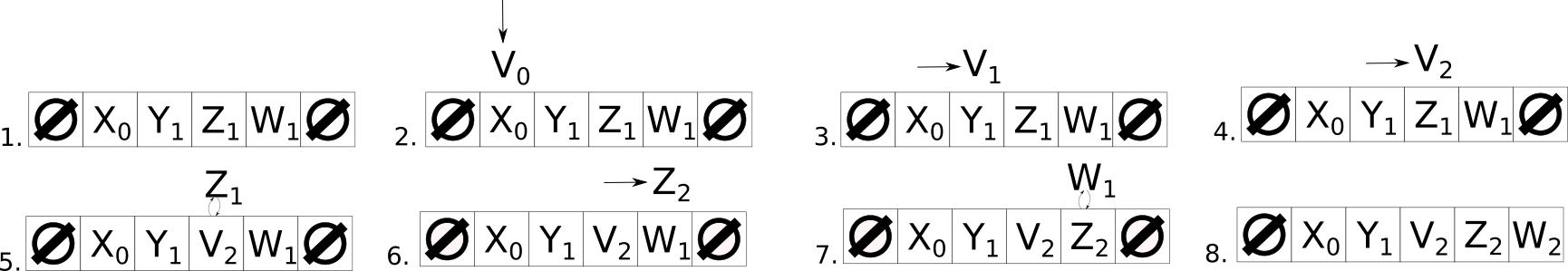}
  \caption{An example Robin Hood insertion where \texttt{V} is inserted into \texttt{X}'s bucket. Each step of the insertion is numbered.}
  \label{fig:RobinHoodInsertion}
\end{figure*}

\begin{figure*}[h]
  \centering
  \includegraphics[width=0.55\textwidth]{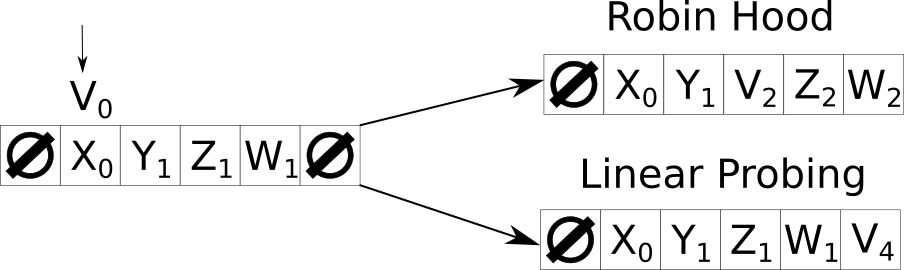}
  \caption{Comparison of Robin Hood to Linear Probing for insertion of \texttt{V}.}
  \label{fig:InsertionComparison}
\end{figure*}

Robin Hood solves this inequality during insertion by moving existing entries around the table. In other words, when the item being relocated has a larger \textit{DFB} than the item currently being examined, these items are swapped, and the search continues for an empty bucket. Once an empty bucket is found, the relocated item is inserted, and the process terminates. Figure \textbf{\ref{fig:RobinHoodInsertion}} shows an example Robin Hood insertion. In all examples the subscripts represent the \textit{DFB} of each entry. Step 1 shows the table initially containing \texttt{X}, \texttt{Y}, \texttt{Z}, and \texttt{W}, each with a \textit{DFB} of 0, 1, 1, and 1 respectively. Step 2 shows that \texttt{V} is to be inserted where \texttt{X} currently resides. Step 3 shows how \texttt{V} doesn't kick out \texttt{X}, as they're equal in \textit{DFB}; the same is true for \texttt{Y} in step 4. Step 5 shows the swap between \texttt{V} and \texttt{Z}, as \texttt{V} is now further away than \texttt{Z} (\textit{DFB} of 2, compared to 1). \texttt{Z} linear probes further down the table in step 6, and in step 7 swaps with \texttt{W}. Finally, in step 8 \texttt{W} lands in the empty bucket at the end, and the insertion process finishes. In summary, \texttt{V} is swapped with \texttt{Z}, \texttt{Z} is swapped with \texttt{W}, and, finally, \texttt{W} is placed in the empty bucket at the end. Figure \textbf{\ref{fig:InsertionComparison}} shows a comparison of the Robin Hood insertion and the same insertion using the Linear Probing collision resolution. As can be seen, the entry \texttt{V} ends up far further away using Linear Probing than using Robin Hood.

\begin{figure*}[ht]
  \centering
  \includegraphics[width=0.95\textwidth]{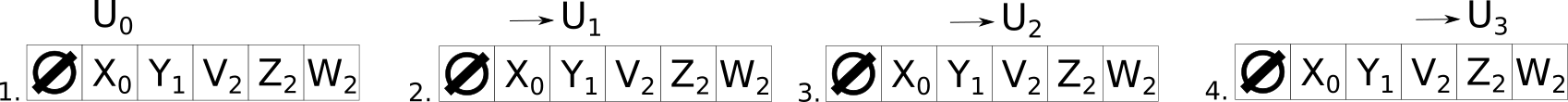}
  \caption{Example Robin Hood search, querying the table for the entry U.}
  \label{fig:RobinHoodSearch}
\end{figure*}

The outcome of this shuffling is a reduced average variance in \textit{DFB} (i.e. reduced probe lengths). Not only does reduced variance make for more predictable and uniform performance, it also allows searches to be culled early, without having to find an empty bucket, which is typically the requirement for calling off a linear probing search. As soon as the item being examined during the search has a lower \textit{DFB} than the current probing \textit{DFB} the search can be called off, with the knowledge that it will be unsuccessful: the item being searched for cannot possibly be present in the table, as it would already have kicked out any items with a lower \textit{DFB} than itself. Figure \textbf{\ref{fig:RobinHoodSearch}} shows an example Robin Hood search operation. The key \texttt{U} is being queried, probing the table as far as the bucket containing \texttt{Z} before terminating. The linear probing is shown in steps 1 to 3, while termination happens at step 4. Termination is possible as \texttt{U} is 3 buckets away from its original bucket, whereas \texttt{Z} is only 2, meaning \texttt{U} can't possibly be in the table. The reasoning is that if \texttt{U} was being inserted it would have displaced \texttt{Z} for having a higher \textit{DFB} than itself; therefore, it cannot possibly be in the table. A similar search taking place using linear probing would have to keep searching until an empty bucket is found, resulting in far more probes at higher load factors, and higher amounts of cache misses too. We refer to this search mechanism as the Robin Hood invariant. Violating this invariant leads to a corrupted table, potentially losing items in the table. These factors enable Robin Hood to support a higher load factor, given that the expected search is only 2.6 probe counts on average for successful searches, and $\mathcal{O}(ln(n))$ for unsuccessful searches \cite{Celis:1986:RHH:13298}.

\begin{figure}[h]
  \centering
  \includegraphics[width=0.8\textwidth]{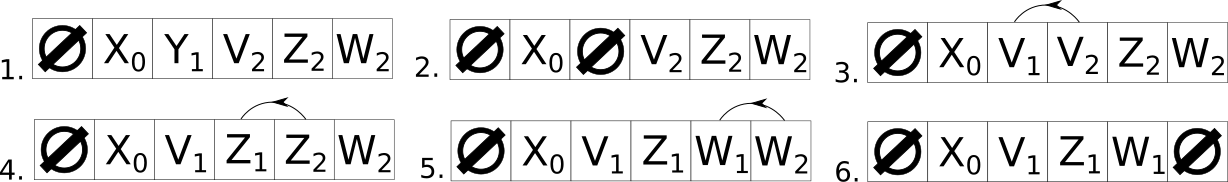}
  \caption{Example Robin Hood deletion. Entry in question is Y.}
  \label{fig:RobinHoodRemoval}
\end{figure}

Deletion in Robin Hood is more complicated than in linear probing. In linear probing one can simply tombstone a bucket, as it does not matter what entry was there before. However, in Robin Hood the \textit{DFB} of the entry matters, as subsequent searches employ this metric to determine if the entry being queried is contained in the table. The solution to this is to either logically mark the entry as deleted, or shift back every entry in front of it until some criterion is met. Logical deletion is unacceptable, as it causes the table to fill up and lose its efficiency, resulting in unnecessary resizes. Backward shifting effectively undoes the insertion of the entry we wish to delete from the table. An example of this is given in Figure \textbf{\ref{fig:RobinHoodRemoval}}. Since the removing thread cannot tell which entries the item to be deleted originally displaced to get there (assuming they aren't at their ideal bucket), we must shift back all items. The shifting is continued until we find an empty bucket or an item in its ideal bucket. Step 1 shows the initial table, step 2 shows the ``nulling'' of \texttt{Y}. Steps 3, 4, and 5 show a backward shifting of entries within the table. Step 6 shows the termination of the shifting, as an empty bucket is ahead of \texttt{W}.

\subsection{K-CAS}
\textit{K-CAS} or, multi-word-compare-and-swap, is an extended version of \textit{CAS}, supporting multiple compare-and-swap operations on many distinct memory locations, all of which either succeed or else fail together. Our algorithm employs a modified version of the \textit{K-CAS} originally proposed by Harris, Kaiser and Pratt \cite{Harris:2002:PMC:645959.676137}. The \textit{K-CAS} algorithm does come at a small cost, reserving an additional 0-2 bits for each word being manipulated by the algorithm. These reserved bits are needed to store run-time type information that allows descriptors to be distinguished from normal values. The standard \textit{K-CAS} interface provides two functions for basic reading and writing, \texttt{K\_CAS\_READ(loc:\ T*) -> T} and \texttt{K\_CAS\_WRITE(loc:\ T*, T val) -> void}, and a mechanism for adding addresses and their values to a descriptor. The rationale for needing dedicated read and write functions is that the values being operated on have specific bits reserved to indicate an ongoing \textit{K-CAS} operation. Both the read and write functions help any pending \textit{K-CAS} operation installed at that particular memory location.

Traditional \textit{K-CAS} implementations require a memory reclaimer system, as each descriptor must be fresh to avoid the ABA problem \cite{ABA_IBM} . However, the specific \textit{K-CAS} implementation we use, developed by Arbel-Raviv and Brown \cite{ArbelRaviv2017ReuseDR}, employs descriptor reuse, thereby eliminating the need for a freshly allocated descriptor for each operation. Given that this implementation does not require a memory allocation per \textit{K-CAS} operation, or a reclaimer, its performance is substantially improved. This enhancement makes \textit{K-CAS} a feasible concurrency primitive and, as we show, allows it to outperform the best lock-based hash table algorithms.

\section{Algorithm}
\subsection{Challenges For Concurrent Robin Hood}
The primary challenge in making Robin Hood concurrent is, unsurprisingly, the modifying operations on the table. Both \texttt{Add} and \texttt{Remove} can modify large parts of the table, with \texttt{Add} potentially performing a global table reorganisation, and \texttt{Remove} potentially shifting back many entries. These problems defeat naive solutions such as \textit{sharded} locks, as \texttt{Add} could potentially grab all of them, leading to no concurrency. Another issue is that of deadlock; insertions at two different points in the table might grab the locks in a cyclic manner, leading to deadlock. For example, the table could have 8 locks with 8 ongoing insertions in 8 different locations. Initially, each insertion will grab one lock corresponding to the original location. If all insertions relocate an entry to another lock section, deadlock occurs. To achieve a truly concurrent implementation of Robin Hood we need an efficient mechanism to update large disparate parts of the underlying table. For this there are two options. The first is \textit{K-CAS}, which became feasible from a performance standpoint thanks to the work of Arbel-Raviv and Brown \cite{ArbelRaviv2017ReuseDR}. The second choice is hardware transactional memory, which we use to provide efficient speculative lock-elision \cite{LockElision}.

\subsection{Overview}
\textit{K-CAS} is a natural choice for relocation-based hash table algorithms. It prohibits a number of issues that typical concurrent algorithms run into, for example, examining whether invariants hold during some intermediate operation or state. \textit{K-CAS} behaves like an expressively weaker transactional memory \cite{HerlihyTransactionalMemory}, but unlike hardware transactional memory \cite{Intel_RTM}, it has well defined progress guarantees. Another reason for using \textit{K-CAS} is that keys can be stored directly in the table, thus improving cache locality. The entry relocations initiated by modifications are summarised into a \textit{K-CAS} descriptor instead of relying on in-place modifications of the table featuring entry relocation information. Threads cannot see a \textit{K-CAS} operation partially completed. Nevertheless, special considerations need to be made for operations when reading the table, as they can experience inconsistent views if applied naively.

\begin{figure*}[ht]
  \centering
  \includegraphics[width=0.95\textwidth]{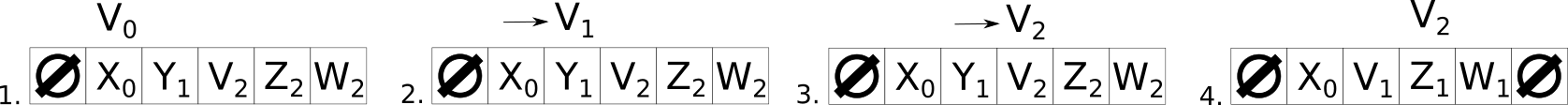}
  \caption{The problem of concurrent searches and removals. The searcher is trying to find \texttt{V} while entry \texttt{Y} is being removed}
  \label{fig:RobinHoodProblem}
\end{figure*}

Since entries can be moved around during concurrent reads, they could inadvertently miss a key due to some ongoing relocation operation. For example, when \texttt{Contains} uses the Robin Hood invariant to terminate a search early, there is a race with a concurrent \texttt{Remove} which could have shifted that particular entry in question back through the table, behind the reader, leading to an incorrect result. This happens when \texttt{Remove} is called on an unrelated entry located in the vicinity of the entry being queried. An example of this phenomenon is illustrated in Figure \textbf{\ref{fig:RobinHoodProblem}}. Here the entry \texttt{V} is being queried while at the same time entry \texttt{Y} is being removed. When the reader gets to the bucket containing \texttt{V}, the remover executes its \textit{K-CAS} operation, shifting a number of entries, including \texttt{V}, backwards. The searcher then checks the bucket after the shift and sees \texttt{Z}, terminates the search, and falsely declares \texttt{V} as not present in the table.

\begin{figure*}[ht]
  \centering
  \includegraphics[width=0.5\textwidth]{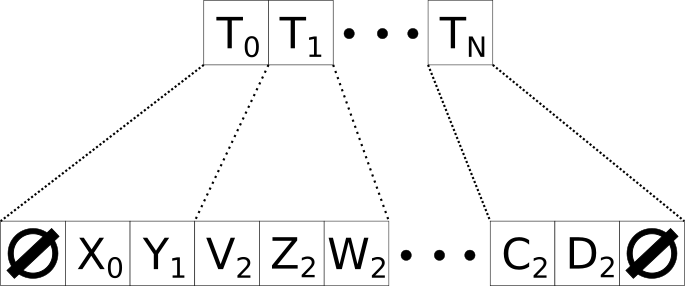}
  \caption{An illustration of how the timestamps are \textit{sharded} across multiple entries.}
  \label{fig:RobinHoodTimestamps}
\end{figure*}

Our solution to avoiding this race is to associate each part of the table with a timestamp, updated upon every relocation. Figure \textbf{\ref{fig:RobinHoodTimestamps}} shows the correspondence of timestamps to physical buckets in the table. A timestamp can be mapped onto several buckets in the table. The mapping of the timestamps is identical to how locks are \textit{sharded} in blocking hash tables like Hopscotch Hashing \cite{Hopper}. When a reader is checking for the presence of a particular key, the reader remembers the timestamps encountered during its search. If the key is found and read atomically, then the search can finish. However, if they key isn't found the reader must check the values of those timestamps after the search has completed. If a discrepancy is found, the search is restarted, otherwise we know for certain the key isn't in the table. When \texttt{Remove} is executing, it increments the timestamp every time it shifts an entry, and similarly for the \texttt{Add} method.

\subsection{Algorithm Methods}
We now outline our algorithm with annotations of code presented in Figures \textbf{\ref{fig:Contains}}, \textbf{\ref{fig:Add}}, and \textbf{\ref{fig:Remove}}, which highlight and explain important parts and what they do. The algorithmic code has been simplified in two areas. The first simplification involves timestamps: the code provided doesn't check if a timestamp has already been added to the list, nor does it check the number of entries per timestamp. The second simplification involves \texttt{Remove} when shuffling elements back. In both cases the code is long but simple, so we have excluded it for the sake of clarity.

\begin{figure}[ht]
\begin{program}
fn Contains(key: K) -> bool \{
  start\_bucket: u64 = hash(key) \% size;
retry:
  timestamps: List<u64> = [];
  for(i = start\_bucket, cur\_dist = 0;; cur\_dist++, i++) \{
    i \%= size;
    // Add the timestamp for that specific index.
    timestamps.append(read\_timestamp(i));
    cur\_key: K = K\_CAS\_load(\&table[i]);
    if(cur\_key == Nil) \{ goto timestamp\_check; \}
    if(cur\_key == key) \{ return true; \}
    distance: u64 = calc\_dist(cur\_key, i); // Robin Hood Invariant
    if (distance < cur\_dist) \{ goto timestamp\_check; \}
  \}
timestamp\_check:
  // Compare every timestamp.
  for(i = start\_bucket, idx = 0; idx < timestamps.size(); i++, idx++) \{
    i \%= size;
    if(list[idx] != read\_timestamp(i)) \{ goto retry; \}
  \}
  return false; // No key
\}
\end{program}
\caption{Pseudo-code for \texttt{Contains}}
\label{fig:Contains}
\end{figure}

\begin{flushleft}
\textit{A - Contains}
\end{flushleft}

All line numbers refer to Figure \textbf{\ref{fig:Contains}}. At the beginning of the \texttt{Contains} method, a timestamp list is created. Line \textbf{8} adds the timestamp for that bucket to the list of timestamps. Line \textbf{9} loads a candidate key from the table; if it's a \texttt{Nil} key, then the search is culled and the timestamps are checked. Line \textbf{11} handles a matching key, returning \texttt{true}. Lines \textbf{12 - 13} handle the case where the distance of the key is too far away from its original bucket to be present in the table. Lines \textbf{17 - 19} check if the timestamp for the particular key has changed since the beginning of the operation, requiring the entire \texttt{Contains} method to be retried, as a key could have been missed.

\begin{figure}[ht]
\begin{program}
fn Add(key: K) -> bool \{
  start\_bucket: u64 =  hash(key) \% size;
retry:
  active\_key: K = key;
  descriptor: K\_CAS\_Desc = create\_descriptor();
  for(i = start\_bucket, active\_dist = 0;; i++, active\_dist++) \{
    i \%= size;
    cur\_timestamp = read\_timestamp(i);
    cur\_key: K = K\_CAS\_load(\&table[i]);
    if(cur\_key == Nil) \{
      descriptor.add(\&table[i], Nil, active\_key);
      res: bool = K\_CAS(descriptor); // Attempt to K-CAS operations
      if(!res) \{ goto retry; \}
      return true;
    \}
    if(cur\_key == key) \{ return false; \}
    distance: u64 = calc\_dist(cur\_key, i); // Robin Hood Invariant
    if (distance < active\_dist) \{
      descriptor.add(\&table[i], cur\_key, active\_key); // Swap keys
      // Increment the timestamp within the descriptor.
      add\_timestamp\_increment(i, cur\_timestamp);
      // Swap active key and kick this key down the table
      active\_key = cur\_key;
      active\_dist = distance;
    \}
  \}
\}
\end{program}
\caption{Pseudo-code for \texttt{Add}}
\label{fig:Add}
\end{figure}

\begin{flushleft}
\textit{B - Add}
\end{flushleft}

All line numbers refer to Figure \textbf{\ref{fig:Add}}. \texttt{Add} is quite similar to the serial version. Lines \textbf{4 - 5} keep track of the entry currently being relocated via the active key, initially setting the variable to the key being inserted. \texttt{Add} also keeps track of the last timestamp bucket it has incremented. This is hidden behind a helper function called \texttt{add\_timestamp\_increment}. Timestamps are incremented to prevent a concurrently running \texttt{Add} or \texttt{Remove} from interfering with the correctness of the method. Lines \textbf{10 - 14} attempt to insert the key currently being relocated into a \texttt{Nil} bucket, retrying the whole  \texttt{Add} operation on failure. Line \textbf{16} returns \texttt{false} upon a key match, as the entry is already in the set. Lines \textbf{17 - 24} check if an entry needs to be relocated, replacing it with the entry currently being relocated in the thread's \textit{K-CAS} descriptor.

\begin{figure}[ht]
\begin{program}
fn Remove(key: K) -> bool \{
  start\_bucket: u64 = hash(key) \% size;
retry:
  timestamps: List<u64> = [];
  descriptor: K\_CAS\_Desc = create\_descriptor();
  for(i = start\_bucket, cur\_dist = 0;; cur\_dist++, i++) \{
    i \%= size;
    // Add the timestamp for that specific index.
    timestamps.append(read\_timestamp(i));
    cur\_key: K = K\_CAS\_load(\&table[i]);
    if(cur\_key == Nil) \{ goto timestamp\_check; \}    
    if(cur\_key == key) \{
      // Shuffle items down until a Nil key or dist(key) == 0
      shuffle\_items(i, cur\_key);
      res: bool = K\_CAS(descriptor);
      if(!res) \{ goto retry; \}
      return true;
    \}
    distance: u64 = calc\_dist(cur\_key, i); // Robin Hood Invariant
    if (distance < cur\_dist) \{ goto timestamp\_check; \}
  \}
timestamp\_check:
  // Compare every timestamp.
  for(i = start\_bucket, idx = 0; idx < timestamps.size(); i++, idx++) \{
    i \%= size;
    if(list[idx] != read\_timestamp(i)) \{ goto retry; \}
  \}
  return false;
\}
\end{program}
\caption{Pseudo-code for \texttt{Remove}}
\label{fig:Remove}
\end{figure}

\begin{flushleft}
\textit{C - Remove}
\end{flushleft}

All line numbers refer to Figure \textbf{\ref{fig:Remove}}. \texttt{Remove} is a combination of \texttt{Contains} and \texttt{Add}. First, \texttt{Remove} tries to find the key. If the key isn't found then, as per \texttt{Contains}, timestamps are checked in case a concurrent \texttt{Remove} or \texttt{Add} has relocated the key during its search. If a key is found, then the process of deletion begins. Lines \textbf{12 - 17} constitute the deletion process. The function \texttt{shuffle\_items} linearly shuffles items back until a \texttt{Nil} key is found, or else an entry with a \textit{DFB} of 0 is found. As mentioned earlier, the function is simple, though expansive, so we exclude it. Each linear shuffle is put into the \textit{K-CAS} descriptor, with line \textbf{15} performing the \textit{K-CAS} operation. The ultimate outcome of this operation is the physical deletion of the entry. Lines \textbf{19 - 20} check for the Robin Hood invariant, terminating the search and checking timestamps if the criterion is met. Like \texttt{Contains}, \texttt{Remove} will restart if there is a discrepancy in the timestamps. Lines \textbf{24 - 26} check the timestamps of each bucket read.

\subsection{Proof Of Correctness}
The proof of correctness is relatively simple and informal. \textit{K-CAS} \cite{Harris:2002:PMC:645959.676137} is itself \textit{linearisable} \cite{Herlihy:1990:LCC:78969.78972}, and \textit{K-CAS} encodes all relocations and timestamp increments to the data structure in its descriptors. An invocation of \textit{K-CAS} essentially turns each modification operation on the table into a transaction. Every \texttt{Add} or \texttt{Remove} call that results in relocations increments a timestamp via the \textit{K-CAS} operation; if any reading thread was to examine the table's contents during a relocation, the reader could compare before and after timestamps so as to ensure that no entry was moved during the search. Furthermore, because each reader also helps the \textit{K-CAS} operation, reading the timestamp will result in one of three outcomes. Either the reader will read a new timestamp value, the same timestamp, or else help the operation complete an ongoing \textit{K-CAS} operation by reading the new timestamp if the operation succeeds or the old one if it fails. Readers need only ensure that the timestamps haven't changed since their initial reading. Since \texttt{Remove} can exit in two ways, it has two different \textit{linearisation} points. The \textit{linearisation} point of \texttt{Add}, and the first code path of \texttt{Remove} are at line \textbf{12} in Figure \textbf{\ref{fig:Add}} and line \textbf{15} in Figure \textbf{\ref{fig:Remove}}, where the \textit{K-CAS} is successfully called. \texttt{Contains} and the second exit point of \texttt{Remove} \textit{linearise} at the point of reading their last timestamp from their search on lines \textbf{8} and \textbf{9} respectively.

\subsection{Progress}
The progress of each operation is parameterised by the progress of the \textit{K-CAS} operation. If the \textit{K-CAS} operation is blocking, then all method calls on the hash table are blocking. However if we have a lock-free implementation of \textit{K-CAS}, then the following classifications for each method occur. Calls to \texttt{Contains} are obstruction-free \cite{ObstructionFree}, as other concurrent operations can cause a relevant timestamp to change, thus forcing the method to restart. Threads calling \texttt{Contains} can starve but they are not blocked if another thread dies. \texttt{Add} has the same progress guarantees as \texttt{Contains} and for the same reasons, other modifying operations can modify relevant timestamps forcing the method to fail and restart. Timestamps are not guaranteed to correspond to the entries the method wants to examine, they are coarse and each timestamp corresponds to a number of entries. \texttt{Remove} must first find the entry it wishes to remove, effectively running the same code as \texttt{Contains} and thus having the same classification. Afterwards, once found, the deletion and subsequent shuffling of that entry is also obstruction-free since other irrelevant operations with the same timestamp can interfere with its progress. In summary, lock-free \textit{K-CAS} calls to \texttt{Contains}, \texttt{Add}, and \texttt{Remove} are obstruction-free.

\section{Performance, Results, and Discussion}
In this section we detail the performance and implementation of our algorithms. All of our code is made freely available online \cite{RobinHoodBenchmark}. This includes \textit{K-CAS} Robin Hood, the hardware transactional \cite{HerlihyTransactionalMemory} variant of Robin Hood Hashing with lock-elision \cite{LockElision}, the implementation of alternative competing algorithms (either coded by us or obtained via online sources), and benchmarking code which allows readers to replicate our results.

\subsection{Experimental Setup}
For our experiments we opted to use a set of microbenchmarks which stressed the hash table under various capacities and workloads. Our benchmarks were run on a 4 CPU machine, with each CPU (Intel\textregistered\ Xeon\textregistered\ CPU E7-8890 v3) featuring 18 cores with 36 hardware threads and 512 GiB RAM. The machine was running Ubuntu 14.04 with kernel version 3.13.0-141. Each thread was pinned to a specific core during testing. When scaling the number of threads, care was taken to pin the thread to a new core, avoiding \textit{HyperThreading\texttrademark} until necessary. Once all non \textit{HyperThreading\texttrademark} cores on each CPU were exhausted, \textit{HyperThreading\texttrademark} was employed thereafter. The rational for this scheduling choice is that algorithms which employ hardware transactional memory are disproportionately penalised by \textit{HyperThreading\texttrademark} and some mainstream operating systems, such as OpenBSD, disable \textit{HyperThreading\texttrademark}. As a result we shed light on how our algorithms would scale on those systems. \textit{NUMA} memory effects were controlled by specifying where each thread could allocate using the \texttt{numactl} command, allocating on the RAM banks closest to the running CPUs as they came into use.

The algorithms used in the experiments are as follows: \textit{K-CAS} Robin Hood, Transactional Lock-Elision Robin Hood Hashing, Hopscotch Hashing \cite{Hopper}, a Lock-Free Linear Probing hash table described by Nielsen and Karlsson \cite{Nielsen:2016:SLH:3016078.2851196}, and Michael's lock-free hash table \cite{Michael2002}.

A number of workload configurations were used in graphing the results. Four load factors of 20\%, 40\%, 60\%, and 80\% were chosen, along with two update workload configurations, namely 10\% and 20\%, referred to herein as ``light'' and ``heavy''. Both of the workload configurations were tried at the specified load factors, and both were used for comparing the different Robin Hood Hashing algorithms against each other, and against competitors. We sized the tables at $2^{23}$ to ensure that they wouldn't fit into the cache, thereby exposing each algorithm's effective cache use. The key space was equal to the size of the table, and was filled to the specified load factors. The scalable JeMalloc \cite{JeMalloc} allocator was used in the experiments and no memory reclamation system was used in algorithms that traditionally require one.

\begin{figure*}
  \centering
  \includegraphics[width=0.8\textwidth]{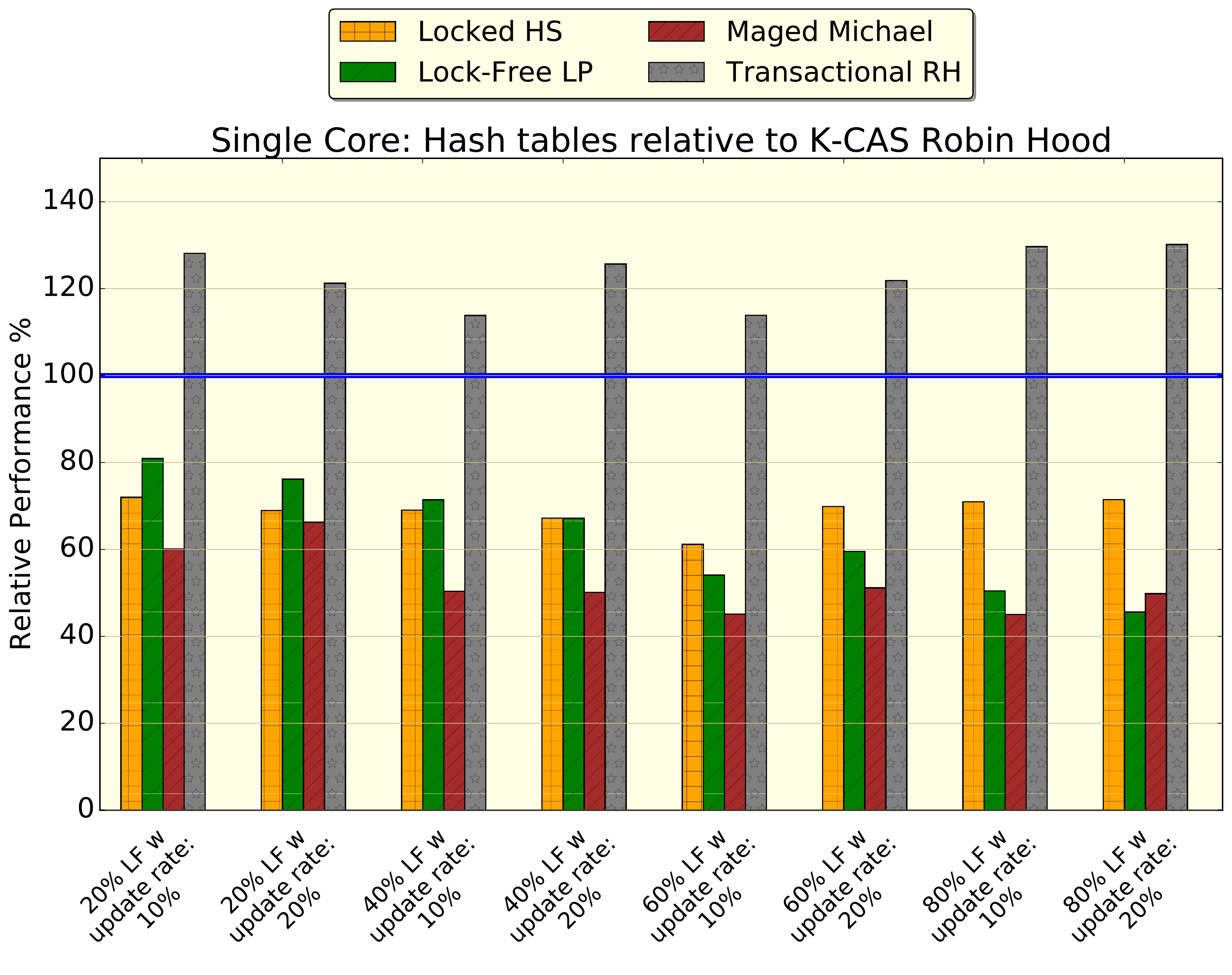}
  \caption{Single-core performance of the hash tables relative to \textit{K-CAS} Robin Hood.}
  \label{fig:RelativePerf}
\end{figure*}

The testing process was carried out as follows: Each thread calls a random method with a random argument from some predefined method and key distribution. All threads are synchronised before execution on the data structure, and for a specified amount of time rather than a specific number of iterations. Each thread counts the number of operations it performed on the structure during the benchmark. The total amount of operations per microsecond for all threads is then graphed. Each experiment was run five times for 10 seconds each, and the average of each result was computed and plotted. All of our algorithms were written in C++11, and compiled with \textit{g++} 4.8.4 with \texttt{O3} level of optimisation. Cache misses were collected via PAPI \cite{PAPI}.

\subsection{Discussion and results}
In order to gain an understanding of general operation overhead, we measured single-core performance. The performance is measured against \textit{K-CAS} Robin Hood. In Figure \textbf{\ref{fig:RelativePerf}} we see that Hopscotch Hashing, Maged Michael's Separate Chaining, and Lock-Free Linear Probing are significantly slower than the other algorithms. The reason for this is two-fold. First, the issue of cache efficiency arises: Lock-Free Linear Probing and separate chaining use dynamic memory allocation, meaning that a pointer dereference is needed for every bucket access. While Hopscotch Hashing does not use dynamically allocated memory, it does put more pressure on the cache by storing the original hash of a key inside the table. However it should be noted that it doesn't put as much pressure on the cache as \textit{K-CAS} Robin Hood, as per Table \textbf{\ref{fig:CacheTable}}. The second issue is the amount of work carried out for every operation: Hopscotch Hashing is the most complicated, executing more code and performing more operations. Transactional Robin Hood has higher performance than \textit{K-CAS} Robin Hood as it does not require to consult the \textit{K-CAS} descriptor and timestamps.

\begin{table}
  \centering
  \begin{tabular}{|l|r|r|r|r|r|r|r|r|}
    \hline
    \multicolumn{1}{|c|}{} & \multicolumn{8}{c|}{Configurations (Load factor w/ Updates)} \\ \hline
  \multicolumn{1}{|c|}{\begin{tabular}[c]{@{}c@{}}Hash\\ Tables\end{tabular}} & \multicolumn{1}{c|}{\begin{tabular}[c]{@{}c@{}}20\%\\ w/\\ 10\%\end{tabular}} & \multicolumn{1}{c|}{\begin{tabular}[c]{@{}c@{}}20\%\\ w/\\ 20\%\end{tabular}} & \multicolumn{1}{c|}{\begin{tabular}[c]{@{}c@{}}40\%\\ w/\\ 10\%\end{tabular}} & \multicolumn{1}{c|}{\begin{tabular}[c]{@{}c@{}}40\%\\ w/\\ 20\%\end{tabular}} & \multicolumn{1}{c|}{\begin{tabular}[c]{@{}c@{}}60\%\\ w/\\ 10\%\end{tabular}} & \multicolumn{1}{c|}{\begin{tabular}[c]{@{}c@{}}60\%\\ w/\\ 20\%\end{tabular}} & \multicolumn{1}{c|}{\begin{tabular}[c]{@{}c@{}}80\%\\ w/\\ 10\%\end{tabular}} & \multicolumn{1}{c|}{\begin{tabular}[c]{@{}c@{}}80\%\\ w/\\ 20\%\end{tabular}} \\ \hline
    Hopscotch Hashing & 88\% & 82\% & 70\% & 71\% & 64\% & 68\% & 59\% & 65\% \\ \hline
    Lock-Free LP & 185\% & 207\% & 227\% & 240\% & 294\% & 305\% & 430\% & 453\% \\ \hline
    Maged Michael & 109\% & 105\% & 95\% & 95\% & 89\% & 93\% & 85\% & 91\% \\ \hline
    Transactional RH & 86\% & 85\% & 82\% & 78\% & 89\% & 86\% & 93\% & 90\% \\ \hline
  \end{tabular}
  \caption{Cache misses relative to \textit{K-CAS} Robin Hood single-core.}
  \label{fig:CacheTable}
\end{table}

The cache results can be seen in Table \textbf{\ref{fig:CacheTable}} as a percentage relative to \textit{K-CAS} Robin Hood for a single-core. These cache statistics were measured over the course of the entire execution of the benchmark for each table and for each configuration. Hopscotch Hashing fairs very well as it is able to skip over irrelevant entries. Lock-Free Linear Probing uses dynamic memory and thus puts enormous pressure on the cache. Another downside is as the table fills up over time with tombstones, it forces operations to take roughly the same amount of time regardless of load factor. This phenomenon is called \textit{contamination} \cite{LinearProbingContamination}. Maged Michael \cite{Michael2002} fairs reasonably well even though it uses dynamic memory. Very few buckets have more than a single node meaning few extra nodes are needed. Transactional Lock-Elision Robin Hood performs better as it does not need to consult an extra timestamp array or any extra \textit{K-CAS} descriptor, which would require an extra level of indirection.

\begin{figure*}
  \centering
  \includegraphics[width=0.45\textwidth]{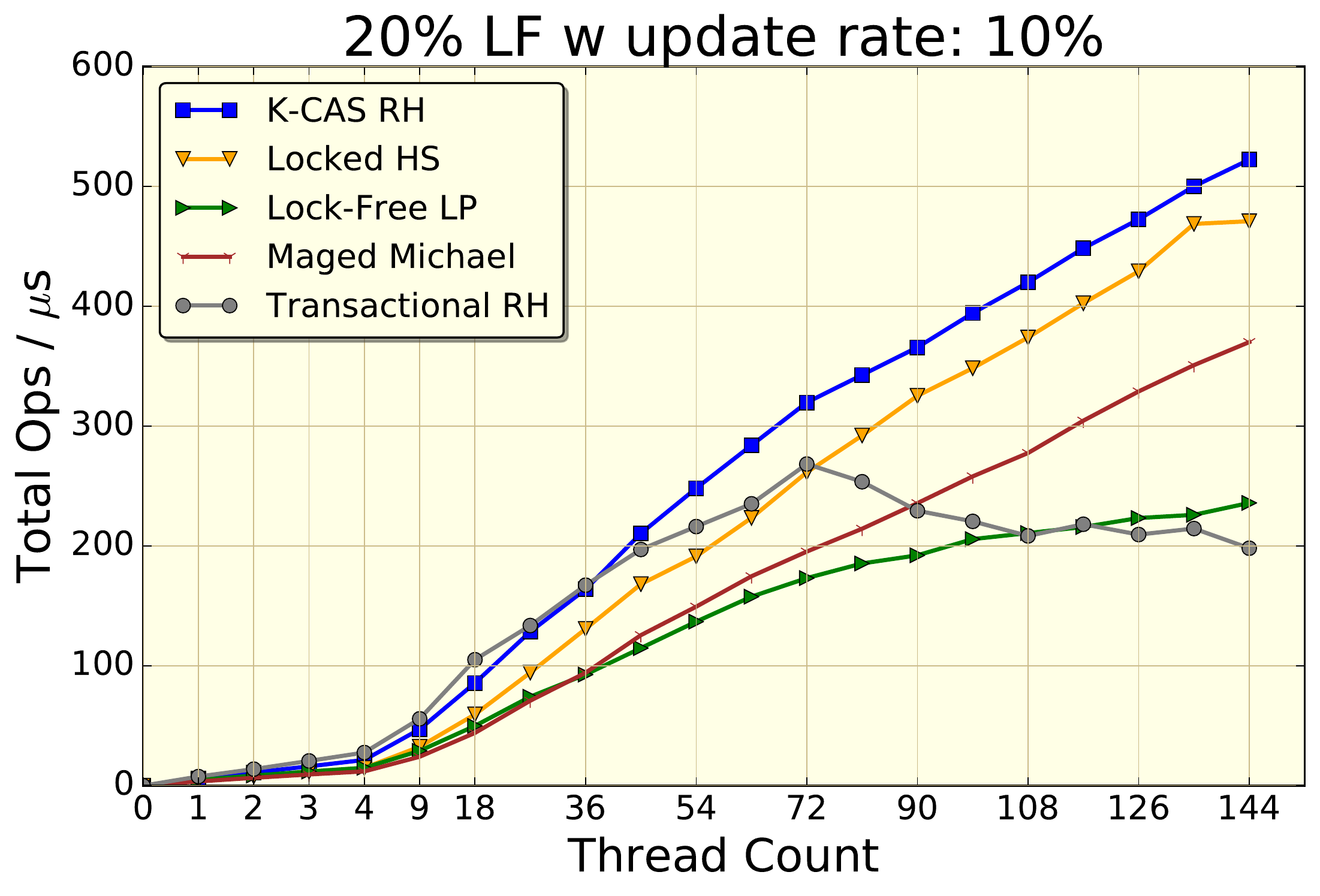}
  \includegraphics[width=0.45\textwidth]{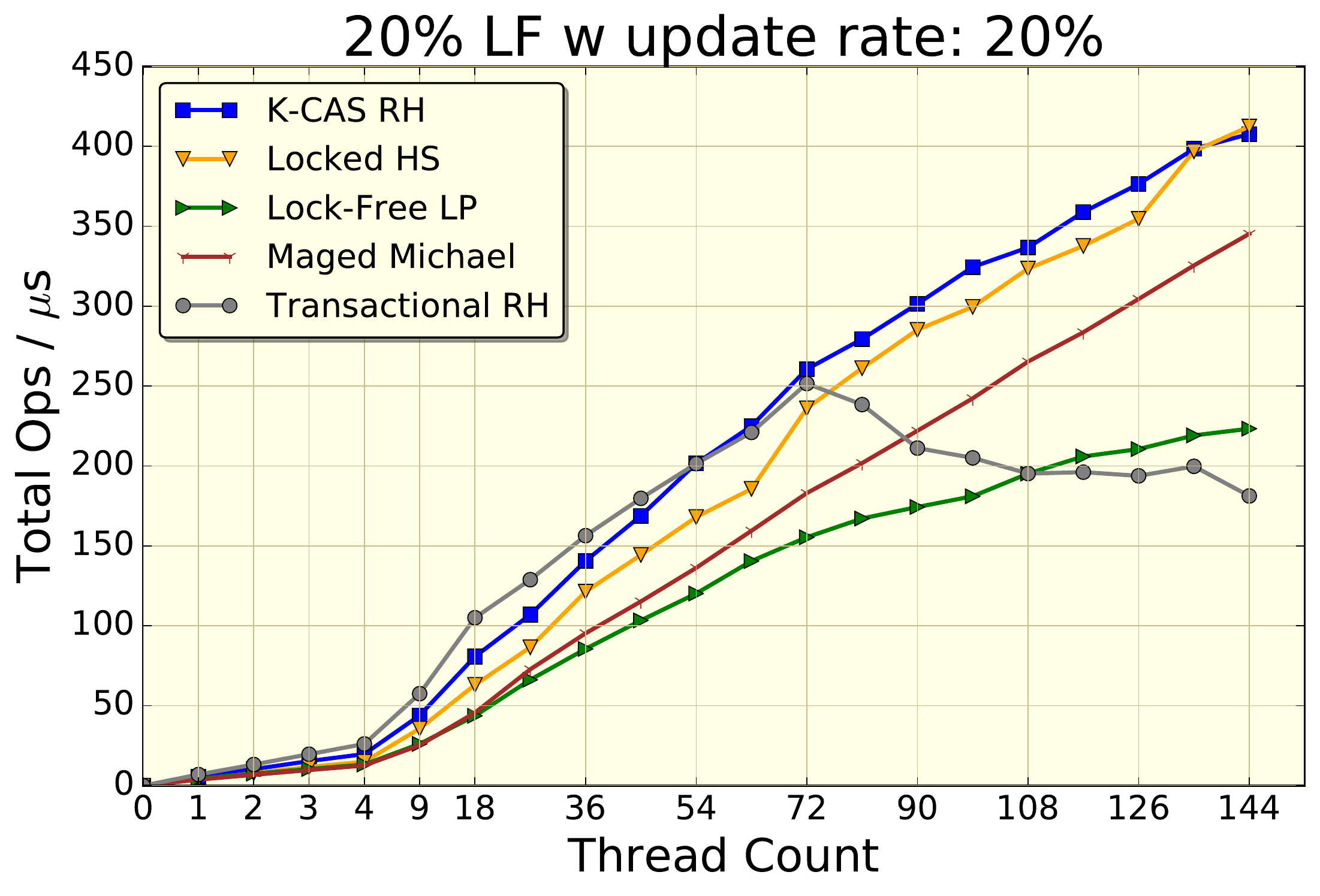}
  \includegraphics[width=0.45\textwidth]{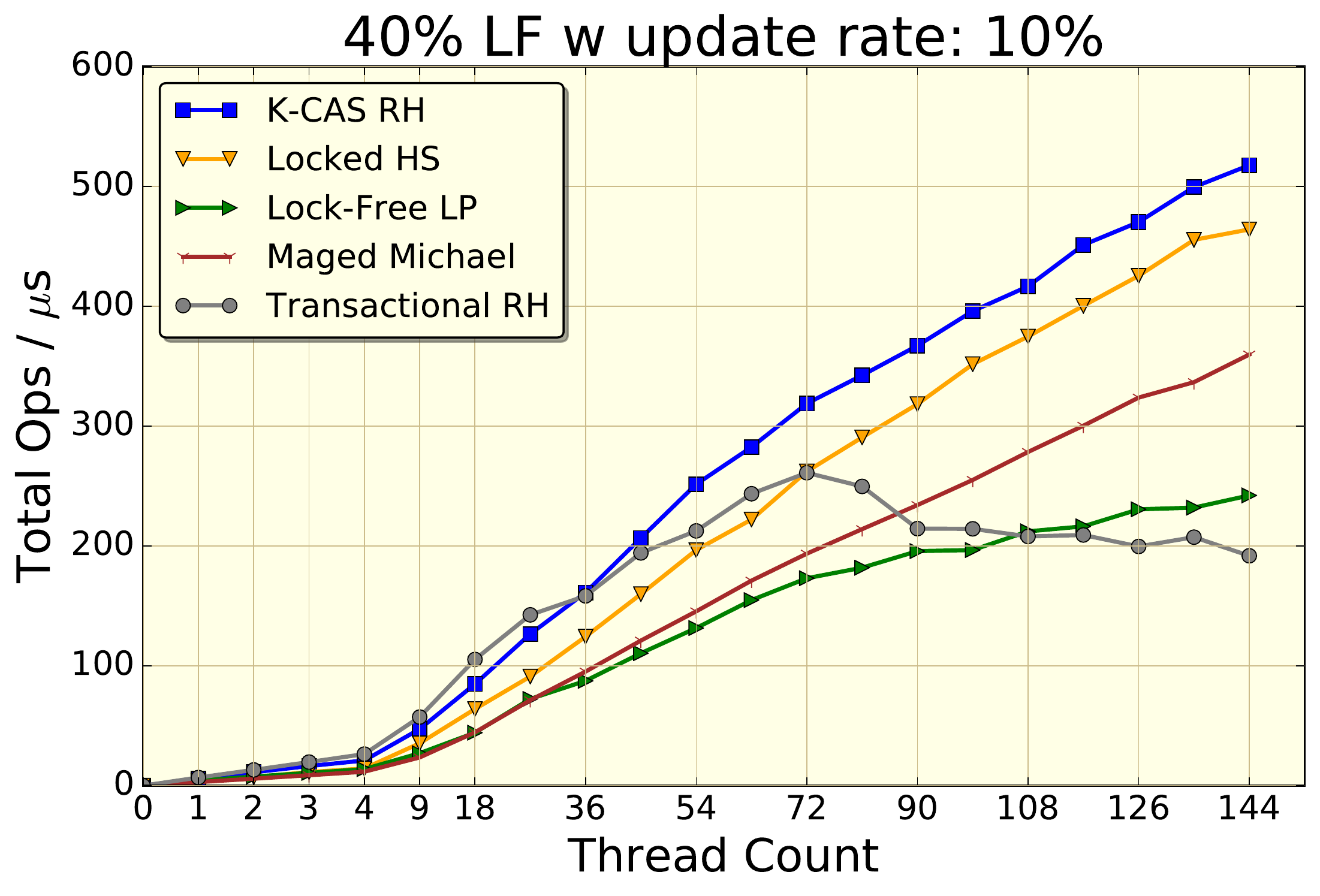}
  \includegraphics[width=0.45\textwidth]{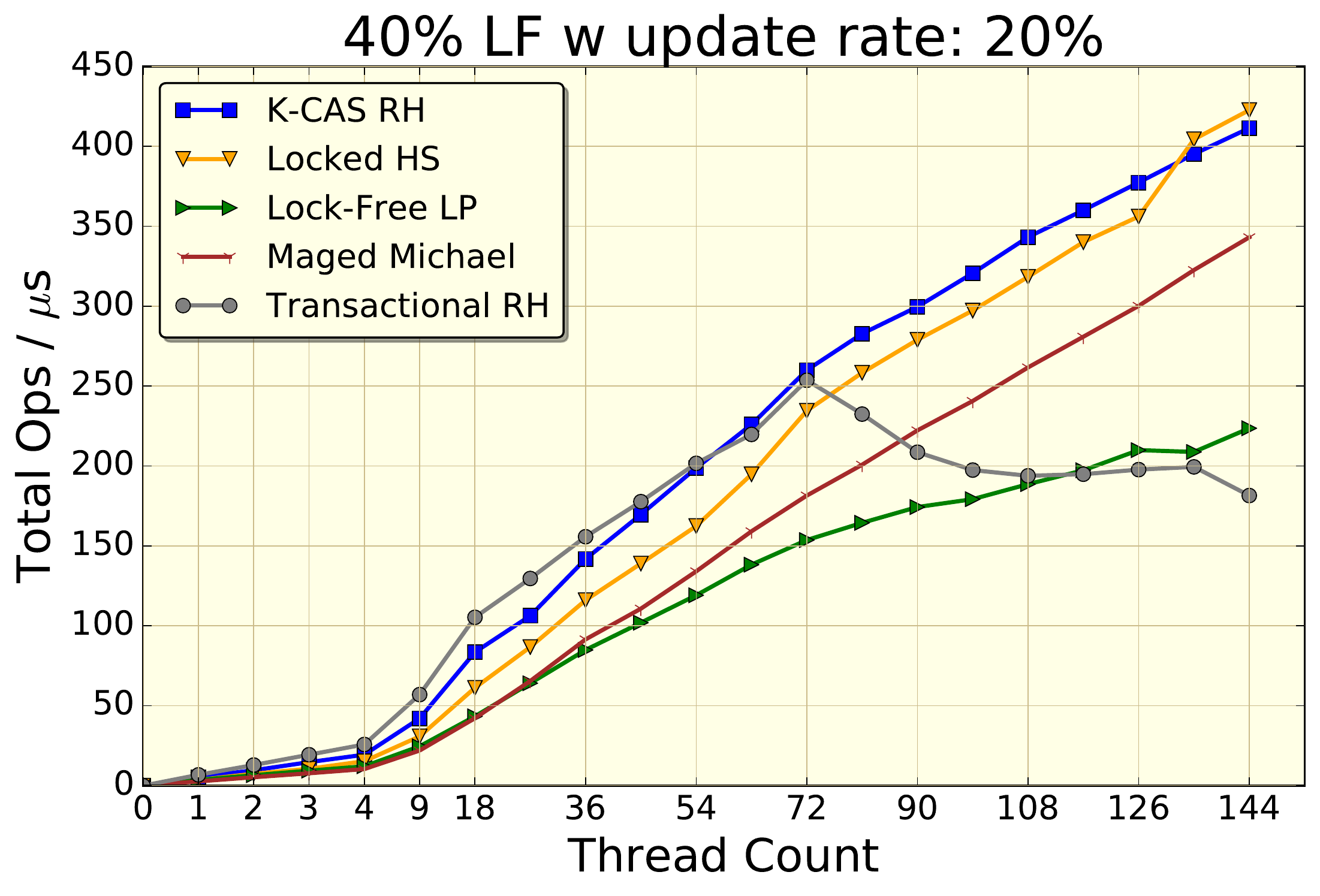}
  \caption{Cumulative operations per microsecond for the hash tables at 20\% and 40\% load factors at two update rates. (Higher is better)}
  \label{fig:NormalLowLoadFactor}
\end{figure*}

\begin{figure*}
  \centering
  \includegraphics[width=0.45\textwidth]{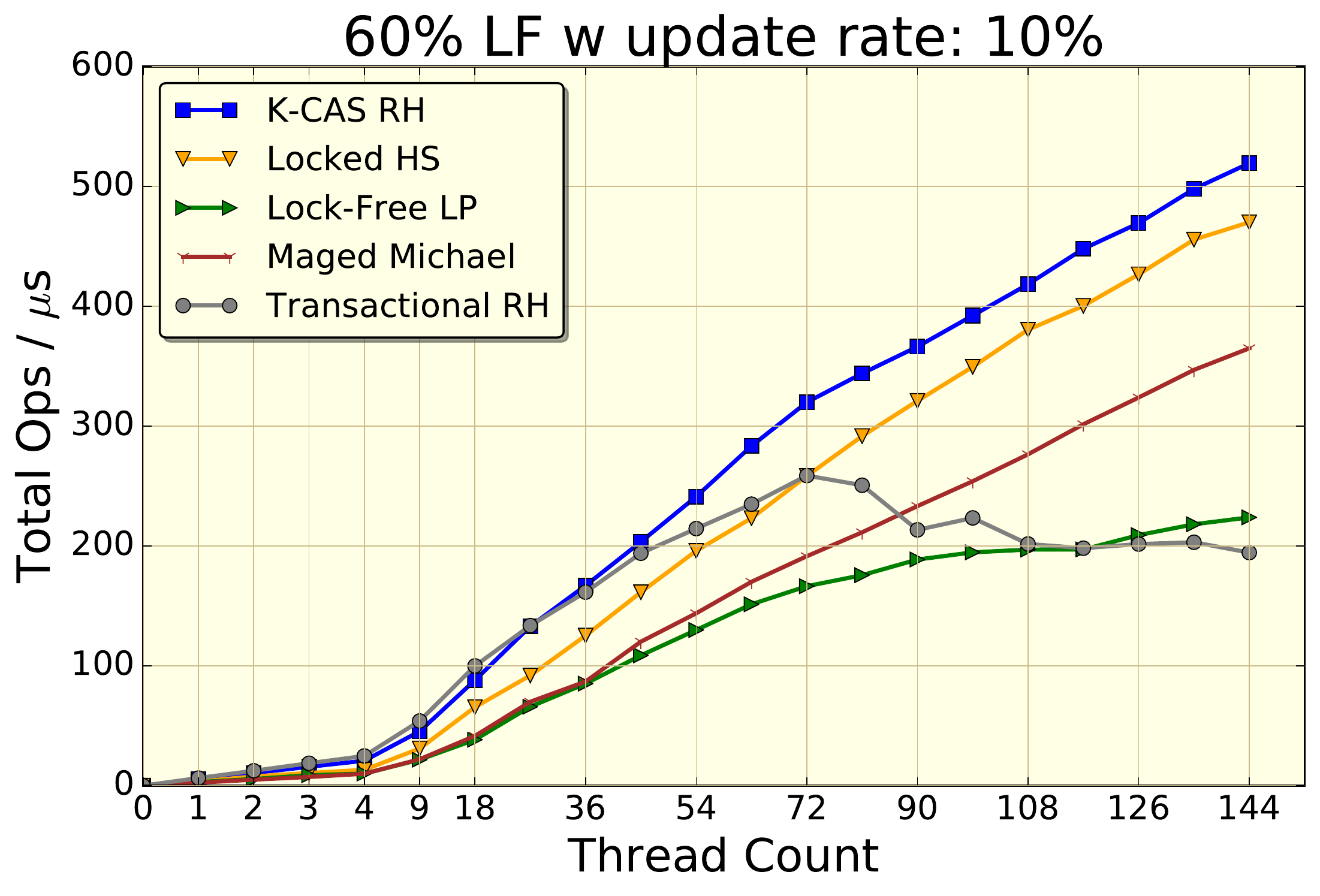}
  \includegraphics[width=0.45\textwidth]{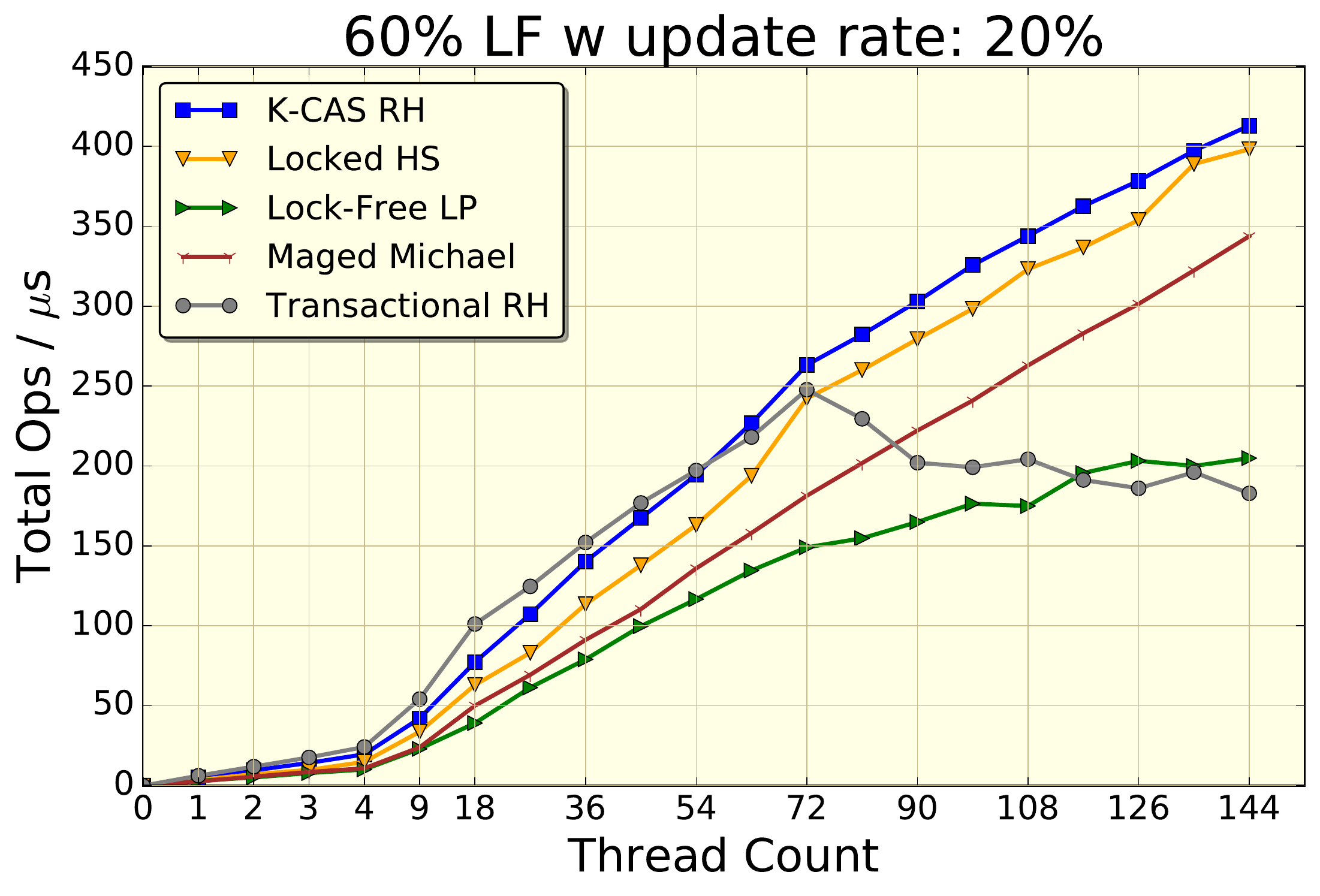}
  \includegraphics[width=0.45\textwidth]{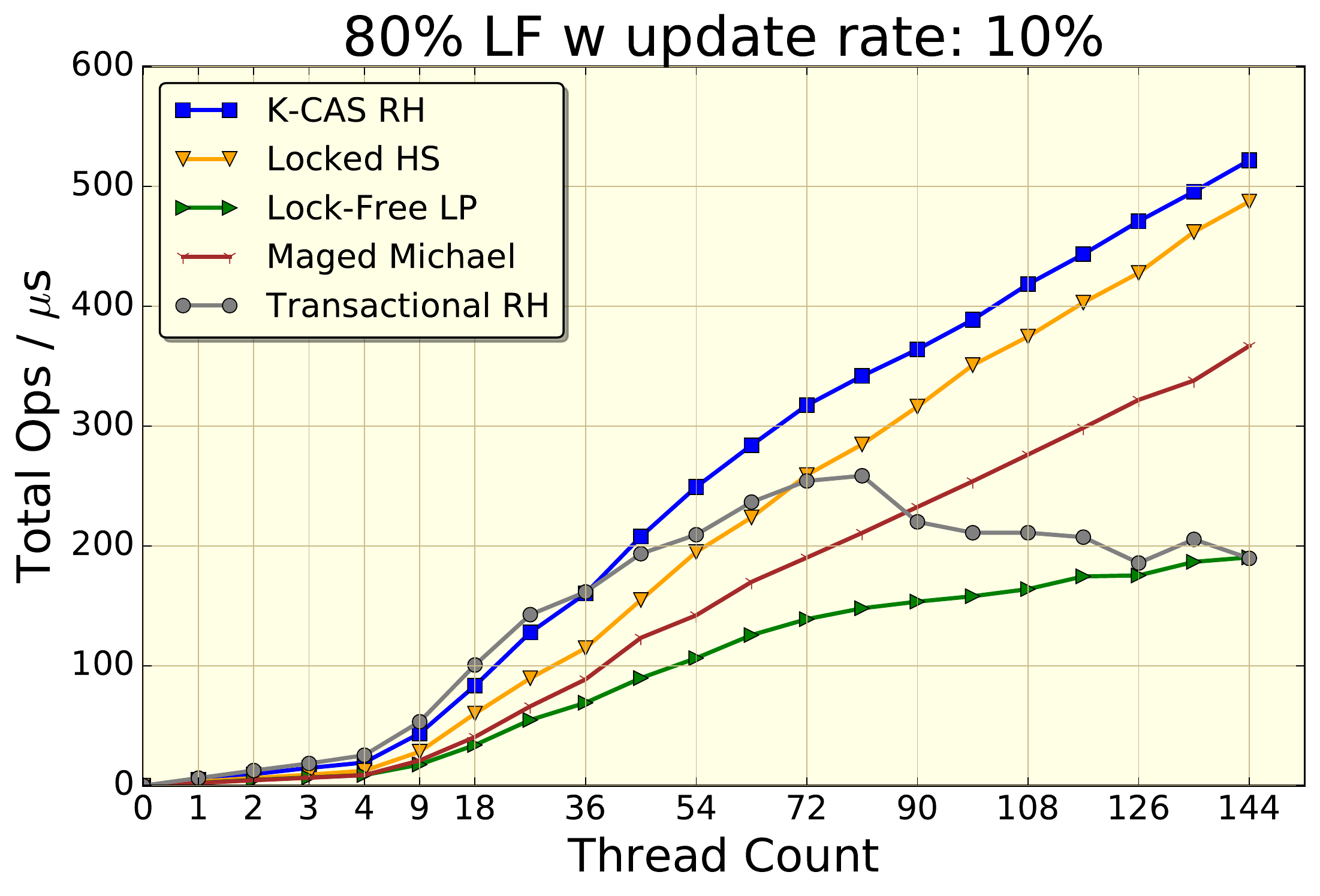}
  \includegraphics[width=0.45\textwidth]{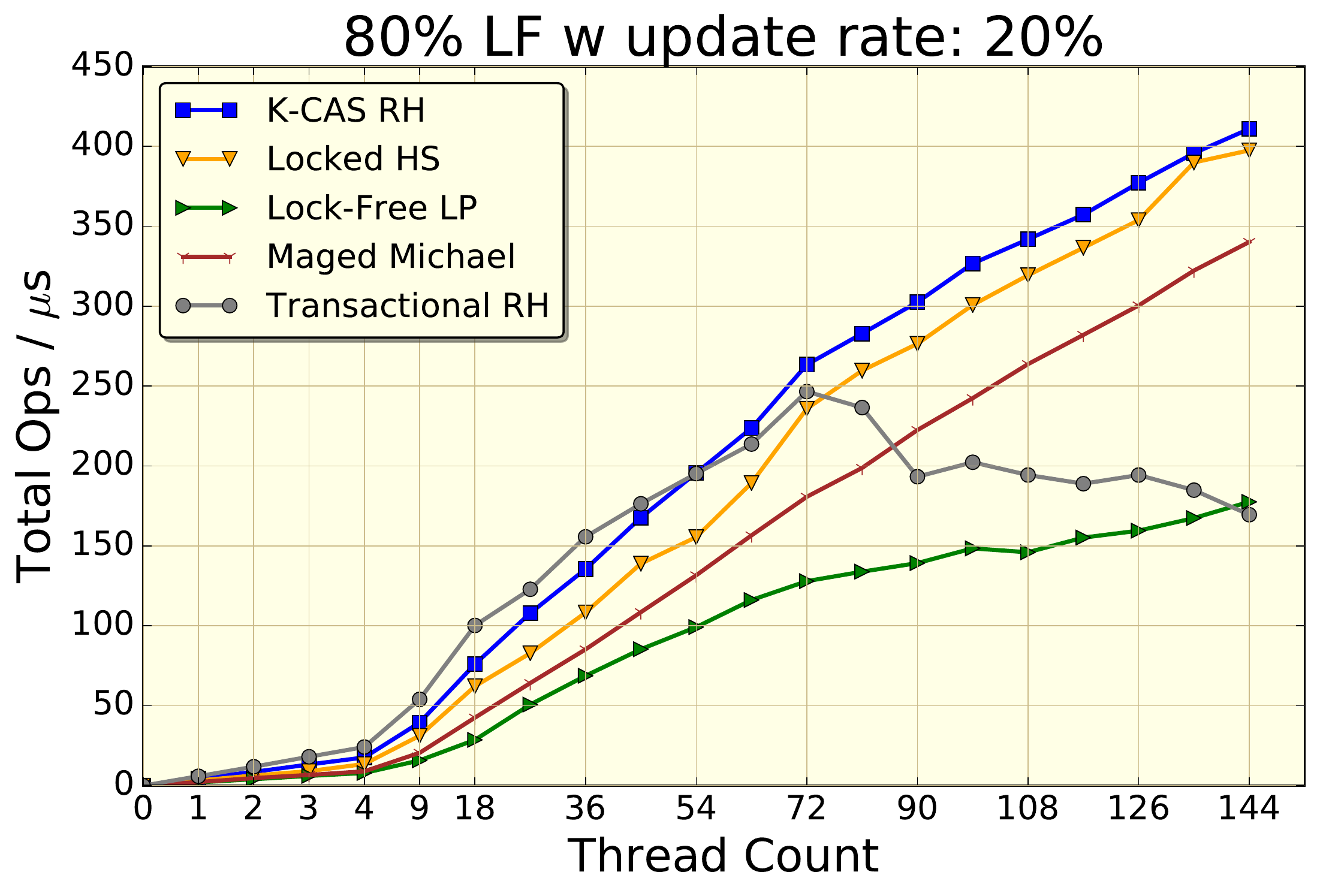}
  \caption{Cumulative operations per microsecond the hash tables at 60\% and 80\% load factors at two update rates. (Higher is better)}
  \label{fig:NormalHighLoadFactor}
\end{figure*}

As is clear from the multi-core results in Figure \textbf{\ref{fig:NormalLowLoadFactor}}, \textbf{\ref{fig:NormalHighLoadFactor}} the \textit{K-CAS} Robin Hood Hashing algorithm either scales better than or is competitive with its opposition. Across all graphs there are two significant dips in performance. The first is at 18 threads, where threads are pinned to a different CPU socket. The use of another socket requires inter-socket communication and \textit{NUMA} effects, reducing overall performance. The second is when increasing from 72 to 81 threads, which is the point where \textit{HyperThreading\texttrademark} kicks in. This kink is particularly pronounced for the transactional Robin Hood variant, which never recovers after that point. Both of these effects become most pronounced when a significant write load is placed on the tables.

Each of the workloads highlights the particular performance characteristics of each hash table. The update-light (10\% writes) workload shows \textit{K-CAS} Robin Hood dominating the competition across all thread counts and load factors. The update-heavy (20\% writes) workload has two distinct outcomes predicated on the table load factor. The lower load factor shows \textit{K-CAS} Robin Hood is slightly beaten out by the Transactional Robin Hood variant until the 72 thread mark, wherein it drops off entirely. \textit{K-CAS} Robin Hood also edges out Hopscotch Hashing throughout most of the update-heavy workload until the very upper end of the thread count, where it is slightly outclassed. However, at higher load factors \textit{K-CAS} Robin Hood demonstrates similar relative performance to the lower load factor benchmarks but maintains its lead over Hopscotch Hashing. All workloads show the gap between \textit{K-CAS} Robin Hood and Hopscotch begins to narrow once \textit{HyperThreading\texttrademark} kicks in. As mentioned earlier, in every configuration the transactional variant of Robin Hood scales very well until 72 threads, after which \textit{HyperThreading\texttrademark} causes a huge kink in performance, and the algorithm never recovers. Maged Michael scales well in the tested workloads, however the gradient of the line isn't steep enough to challenge \textit{K-CAS} Robin Hood or Hopscotch Hashing. Lock-Free Linear Probing does the worst of all, ending up at the same end point as Transactional Robin Hood.

\subsection{Future Work}
An issue we don't deal with is \textit{resize}, specifically, when to resize the table and how to do it. Lock-free \textit{resize} methods have been discussed in the literature (\cite{CliffClick}, \cite{Shalev03}). To the best of our knowledge there has not yet been a formally published generic hash table resize method. Another item for future work is a combination of lock-free \textit{K-CAS} and transactional \cite{HerlihyTransactionalMemory} memory, such as the algorithm presented by Trevor Brown \cite{BrownTrees} for lock-free trees. A similar exploration for Robin Hood and other the hash tables benchmarked would be of interest. Work done by Siakavaras et el. \cite{MassiveRBTreesTM} shows that naive application of hardware transactional systems to data-structures typically perform poorly and require algorithm modifications to operate efficiently. Future work aimed at determining the optimal modifications for Robin Hood might elevate its performance for hardware transactional memory.

\section{Conclusion}
We have presented an obstruction-free Robin Hood Hashing algorithm which achieves superior performance relative to other concurrent algorithms with similar capabilities, and a hardware transaction variant which demonstrates best in class performance. These results establish that the Robin Hood algorithm is algorithmically suited to concurrency. Our experiments have shown that it scales strongly at all thread counts. Unlike linear probing, the algorithm can scale effectively to significantly higher load factors. It is very simple in nature, relying primarily on an efficient \textit{K-CAS}, as well as being highly portable, using only single word compare-and-swap instructions, with no memory reclaimer required. Our results improve upon the state of the art in the field which is more than 10 years old.


\bibliography{opodis2018}

\end{document}